\documentclass[12pt]{article}
\usepackage{amssymb}
\usepackage{amsmath,,calrsfs}
\usepackage{slashed}

\usepackage[dvips]{epsfig}

\newcommand{\be}{\begin{equation}}
\newcommand{\ee}{\end{equation}}
\def\bea{\begin{eqnarray}}
\def\eea{\end{eqnarray}}

\newcommand{\bn}{\begin{eqnarray}}
\newcommand{\en}{\end{eqnarray}}

\newcommand{\p}{\partial}

\newcommand{\nn}{\nonumber}

\newcommand{\no}{\noindent}

\def\bea{\begin{eqnarray}}
\def\eea{\end{eqnarray}}

\newcommand{\beq}{\begin{eqnarray}}
\newcommand{\eeq}{\end{eqnarray}}
\begin{document}

\title{\textbf{Duality and unitarity of massive spin-3/2 models in $D=2+1$ }}
\author{E. L. Mendon\c ca \footnote{eliasleite@feg.unesp.br}, H.L. de Oliveira \footnote{helderluiz10@gmail.com} , P.H.F. Nogueira\footnote{pedro.fusco@unesp.br }\\
\textit{{UNESP - Campus de Guaratinguet\'a - DFQ} }\\
\textit{{Av. Dr. Ariberto Pereira da Cunha, 333} }\\
\textit{{CEP 12516-410 - Guaratinguet\'a - SP - Brazil.} }\\
}
\date{\today}
\maketitle

\begin{abstract}
In this work we provide a triple master action interpolating among three self-dual descriptions of massive spin-3/2 particles in $D=2+1$ dimensions. Such result generalizes a master action previously suggested in the literature. We also show that, surprisingly a shorthand notation in terms of differential operators applied in the bosonic cases of spins 2 and 3, can also be defined to the fermionic case. With the help of projection operators, we have also obtained the propagator and analyzed unitarity in $D$ dimensions of a second order spin-3/2 doublet model. Once we demonstrate that this doublet model is free of ghosts, we provide a master action interpolating such model with a fourth order theory which has several similarities with the spin-2 linearized New Massive Gravity theory.
\end{abstract}
\newpage

\section{Introduction}

The seminal work by Deser, Jackiw and Templeton \cite{djt} provides a dynamically and topologically nontrivial theory for gravity in three dimensions. The model describes a single massive spin-2 excitation, and even being of third order in derivatives it is free of ghosts. It is usually called the Topologically Massive Gravity TMG. It is remarkable that in the linearized level one can demonstrate that, through the generalized soldering procedure \cite{soldering2} one can joint opposite helicities $+2$ and $-2$ in a unique doublet model, which consists of the linearized version of the so called New Massive Gravity NMG \cite{BHT}.

Supersymetric extensions of TMG and NMG also exists \cite{Andringa} and the fermionic actions are closely related to the bosonic ones regarding their invariances and order in derivatives. The introduction of a field strenght $f_{\mu}(\psi)=\epsilon_{\mu\nu\alpha}\p^{\nu}\psi^{\alpha}$, where $\psi^{\alpha}$ is a two componet Majorana vector-spinor, allow us the use of the dynamically trivial first order Chern-Simons like term, which is called the Rarita-Schwinger term $\epsilon_{\mu\nu\alpha}\bar{\psi}^{\mu}\p^{\nu}\psi^{\alpha}$ in three dimensions. One can demonstrate that, starting with the the first order in derivatives model suggested in \cite{deser3/2} one can interpolate it with a second order model given by \cite{deserkay}. Such procedure is possible thanks to the construction of a master action. Here, we suggest a generalization of such master action, which consists of a triple master action, interpolating the first two models with a third order in derivatives model also describing the single massive spin-3/2 particle in $D=2+1$. This third order action is precisely the fermionic part of the suspersymetric extension of TMG obtained by \cite{Andringa}.

Those models describing single excitations will be called the singlet models, and we will reffer to it as self-dual descriptions using the symbol $SD(i)$, where $(i=1,2,3)$ indicates the order in derivatives of the given model . We have verified that \cite{soldering32}, through the soldering procedure one can tie together different helicities $+3/2$ and $-3/2$ obtaining doublet models describing both helicities. Such models have several similarities with the bosonic companions of spin-2, such as the Fierz-Pauli theory and the NMG, both in the linearized version. Here we provide a master action interpolating between second order and fourth order doublet models, which will be identified by the symbol $D(i)$ with $(i=2,3)$ also indicating order in derivatives. In order to demonstrate that the second order model is free of ghost we also provide its propagator and study its unitarity showing that the second order model is free of ghosts in $D$ dimensions.

\section{Introducing  a shorthand notation and the models}
The spin-3/2 massive particle will be described in terms of the Majorana two component vector-spinor $\psi_{\mu}$, where the greek indices represents space-time coordinates while the spinorial indices has been suppressed for sake of simplicity. In $D=2+1$ dimensions, such field has then six independent components. All the theories studied here will be free of interactions with other fields, but sources $j_{\mu}$ will be considered in order to study unitarity and to compare correlation functions. The signature used along the text will be mostly plus $(-,+,+)$, while the gamma matrices are indeed Pauli matrices given by: $\gamma^0=i\sigma_2, \gamma^1=\sigma_1$ and $\gamma^2=\sigma_3$. It is useful to keep in hand the following identities $\gamma_{\mu}\gamma_{\nu}=\eta_{\mu\nu}+\epsilon_{\mu\nu\alpha}\gamma^{\alpha}$ , $\left\lbrace\gamma_{\mu},\gamma_{\nu}\right\rbrace= 2\eta_{\mu\nu}$ and $\left\lbrack\gamma_{\mu},\gamma_{\nu}\right\rbrack= 2 \epsilon_{\mu\nu\alpha}\gamma^{\alpha}$. 

In order to deal with the master action approach, it is useful to define a shorthand notation. Here, we use a quite similar idea to \cite{dualdesc}, where:

\bea \int d^3x\, \epsilon^{\mu\nu\alpha} \bar{\psi}_{\mu}\gamma_{\nu}\psi_{\alpha}&\equiv& \int (\bar{\psi}\psi) ,\label{mt}\\
\int d^3x\, \epsilon^{\mu\nu\alpha} \bar{\psi}_{\mu}\p_{\nu}\psi_{\alpha}&\equiv& \int \bar{\psi}\,  \cdot d \psi .\label{cst}\eea

\no The term (\ref{mt}) will be used exclusively as a mass term. The first order in derivatives term is a Chern-Simons like term. In the literature it is called the Rarita-Schwinger action in three dimensions. Similarly to its spin-1 counterpart it is gauge invariant under reparametrizations of the form $\delta_{\Lambda} \psi_{\mu}=\p_{\mu}\Lambda$. The second order term is defined through the  ``field strength'' tensor $f^{\mu}(\psi)=\epsilon^{\mu\nu\alpha}\p_{\nu}\psi_{\alpha}$ as:

\bea \int d^3x\, \bar{f}^{\mu}(\psi)\gamma_{\nu}\gamma_{\mu}f^{\nu}(\psi)= \int d^3x\, \epsilon^{\gamma\beta\mu}\bar{\psi}_{\gamma}\p_{\beta}\Omega_{\mu}(\psi) \equiv \int \bar{\psi} \cdot d \Omega(\psi).\eea

\no We notice that due to the Bianchi identity $\p_{\mu}f^{\mu}(\psi)=0$ the second order action is also gauge invariant under reparametrizations. Additionally, we have defined the useful operator $\Omega_{\mu}(\psi)=\gamma_{\nu}\gamma_{\mu}f^{\nu}(\psi)$.  It is remarkable that once one define the $\Omega$ operator of something, and we can deal with a structure of derivatives a la Chern-Simons, we mean $\epsilon^{\mu\nu\alpha}\p_{\nu}= (\cdot d)^{\mu\alpha}$, we can write all the different models in the shorthand notation. This has been working so far also for different spins, see for example the spin-2 and spin-3 cases \cite{dualdesc, mas3}.

In the next sections we suggest master actions interpolating among self-dual descriptions and doublet models. The self-dual models describe a unique massive particle of helicity $+3/2$ {\bf or} $-3/2$, which we have been calling a singlet. They differ each other by an order in derivatives and by its invariance under local transformations. In terms of the shorthand notation we write them as:

\be S_{SD(1)}= \int \,\left[ -\bar{\psi}\,\cdot d \psi + m \, (\bar{\psi}\, \psi)\right],\label{sd1}\ee

\no where the subscript in the action indicates that this is a self-dual description of first order in derivatives, as also will happen for the other models. The action (\ref{sd1}) was first introduced in \cite{deserkay}. In the present case the model has no gauge invariance because of the presence of the mass term. By imposing gauge invariance to this model we demonstrate \cite{nge32} that one can obtain the second order self-dual  model \cite{deser3/2} which is given by:

\be S_{SD(2)}= \int \,\left[ -\frac{1}{2m}\bar{\psi}\,\cdot d \Omega(\psi) + \bar{\psi}\,\cdot d \psi \right].\label{sd2}\ee

\no Now the model is invariant under reparametrizations of the form $\delta \psi_{\mu}=\p_{\mu}\Lambda$. The second order term plays the role of the Maxwell term if one thinks in a comparison with the spin-1 case or the linearized  Einstein-Hilbert term of the spin-2 field theories. Once the second order term is invariant under gamma traceless transformations of the form $\delta_{\xi} \psi_{\mu}= \gamma_{\mu} \xi$ while the first order term breaks it, in \cite{nge32}, we could again impose such gauge transformation obtaining the third order self-dual model given by:

\be S_{SD(3)}= \int \,\left[-\frac{1}{4m^2}\bar{\Omega}(\psi)\,\cdot d \Omega(\psi)  + \frac{1}{2m}\bar{\psi}\,\cdot d \Omega(\psi)\right].\label{sd3}\ee 

\no The third order term could be compared to the third order linearized topologically Chern-Simons term of the spin-2 theories. Such model is precisely the fermionic part of the Topologically Massive Supergravity theory of \cite{Andringa}. Notice that a new kind of term has appeared and once we know how the shorthand notation works it is straightforward to see that $\bar{\Omega}\cdot d \Omega=\epsilon^{\mu\nu\alpha}\bar{\Omega}_{\mu} \p_{\nu} \Omega_{\alpha}$. 

In \cite{soldering32}, we have demonstrated that by soldering self-dual models with different helicities and masses, we end up with doublet models. Such models support both helicities in its descriptions i.e.: $+3/2$ {\bf and} $-3/2$. In the case of equal masses the result coming from the soldering of two $SD(2)$ models is given by:

\be S_{D(2)}= \int \,\left[ -\frac{1}{4}\bar{\psi}\,\cdot d \Omega(\psi) + \frac{m^2}{2} \, (\bar{\psi}\, \psi) \right].\label{d2}\ee

\no Where we have set the coefficients in front of terms in such a way that we can reduce the equations of motion to the Kein-Gordon equation. Here the subscript in the action  is meaning that we have a doublet description of second order in derivatives. 

Finally, by soldering two third order self-dual models one can obtain a fourth order doublet model, which we claim to be the spin-3/2 analog example of the spin-2 New Massive Gravity in its linearized version.   In the shorthand notation
we can write it as:
\be S_{D(4)}= \int \,\left[ \frac{1}{16m^2}\bar{\Omega}\,\cdot d \Omega(\Omega) +\frac{1}{4}\bar{\psi}\,\cdot d \Omega(\psi)  \right].\label{d4}\ee

We could add to this discussion the fact that, the doublet models can also be connected through symmetry arguments. It is easy to see that (\ref{d2}) has no gauge invariance under reparametrizations due to the presence of the mass term. By implementing such symmetry we get the fourth order action, invariant under reparametrizations and ``gamma-Weyl'' transformations as also happens with NMG.

In the next section we add source terms to the lagrangians in order to study unitarity as well as to compare correlation functions. The source term will be automatically coupled to gauge invariant combinations when this is necessary.

\section{Unitarity of the second doulblet model}

The particle content of the second order doublet model can be investigated at the classical level through its equations of motion. However, we do not have a proof of unitarity, by analyzing the propagator of the theory at tree level. Here we use a spin-3/2 projection operator basis to access the particle content of the model and then we saturate the propagator between conserved sources. Just for sake of simplicity we multiply the lagrangian (\ref{d2}) by a numerical factor $4$. The second order doublet model given by (\ref{d2}) can then be explicitly written as:
\bea
\mathcal{L} &=& \partial^\theta \overline{\psi}^\lambda \gamma^\mu \gamma_\mu \partial_\theta \psi_\lambda + \partial^\lambda \overline{\psi}^\nu \gamma_\nu \gamma^\theta \partial_\theta \psi_\lambda + \partial^\nu \overline{\psi}^\theta  \gamma_\nu \gamma^\lambda \partial_\theta \psi_\lambda - \partial^\theta \overline{\psi}^\nu \gamma_\nu \gamma^\lambda \partial_\theta \psi_\lambda \nn \\ 
&-& \partial^\nu \overline{\psi}^\lambda \gamma_\nu \gamma^\theta \partial_\theta \psi_\lambda - \partial^\lambda \overline{\psi}^\theta \gamma^\mu \gamma_\mu \partial_\theta \psi_\lambda - 2m^2 \overline{\psi}_\mu   \gamma^\mu \gamma^\alpha \psi_\alpha + 2m^2\overline{\psi}_\mu\psi^\mu
\label{9}\eea

\no Notice that, once we work with the derivative term of (\ref{d2}) and also use the identity $\gamma_{\mu}\gamma_{\nu}=\eta_{\mu\nu}+\epsilon_{\mu\nu\alpha}\gamma^{\alpha}$ in the mass term, the lagrangian becomes independent of the dimension. We would like to write the lagrangian (\ref{9}) in a bilinear form using the spin projection and transition operators in $D$ dimensions given by \cite{bernard} for example. Then we have the projectors:

\bea
\left(P^{(3/2)}\right)_{\mu\nu} &=& \eta_{\mu\nu} - \frac{1}{D-1}\gamma_\mu \gamma_\nu - \frac{1}{p^2(D-1)}(\slashed{p}\gamma_\mu p_\nu + p_\mu \gamma_\nu\slashed{p}) + \frac{p_\mu p_\nu}{2p^2}, \\
\left(P^{(1/2)}_{11}\right)_{\mu\nu} &=& - \frac{3}{D-1} \frac{p_\mu p_\nu}{p^2} + \frac{1}{D-1}\gamma_\mu \gamma_\nu + \frac{1}{p^2(D-1)}(\slashed{p}\gamma_\mu p_\nu + p_\mu \gamma_\nu \slashed{p}), \\
\left(P^{(1/2)}_{22}\right)_{\mu\nu} &=& \frac{p_\mu p_\nu}{p^2},
\eea

\no and the transition operators:

\bea
\left(P^{(1/2)}_{12}\right)_{\mu\nu} &=& \frac{1}{p^2\sqrt{D-1}}(p_\mu p_\nu - \slashed{p} p_\nu \gamma_\mu), \\
\left(P^{(1/2)}_{21}\right)_{\mu\nu} &=& \frac{1}{p^2\sqrt{D-1}}(\slashed{p} p_\mu \gamma_\nu - p_\mu p_\nu).
\eea

\no One can verify that they obey the orthogonality relations: 

\be
\left(P^I_{ij}\right)_{\mu\nu}\left(P^J_{kl}\right)^{\nu\rho} = \delta_{IJ}\delta_{jk}\left(P^I_{il}\right)^\rho_\mu\quad,\quad (I,J)=1,2, \quad,\quad (i,j,k,l)= 1,2.
\ee

Once the lagrangian is put in the bilinear form, one can invert the sandwiched operator using the spin-projection basis and find the following propagator in the space of derivatives:

\bea
G^{-1}_{\mu\nu} &=& \frac{1}{D\Box - \slashed{\partial}^2 - 2m^2}\left(P^{(3/2)}\right)_{\mu\nu} + \frac{\Box - \slashed{\partial}^2}{K}\left(P^{(1/2)}_{11}\right)_{\mu\nu}\nn \\ 
&+& \frac{\Box - \slashed{\partial}^2 + 2m^2(D - 2)}{K}\left(P^{(1/2)}_{22}\right)_{\mu\nu} - \frac{2m^2\sqrt{D-1}}{K}\left(P^{(1/2)}_{12} + P^{(1/2)}_{21}\right)_{\mu\nu},
\eea
where $G^{-1}$ is the inverse of the sandwiched operator and  $K = (\Box - \slashed{\partial}^2)^2 + 2m^2(D-2)(\Box-\slashed{\partial}^2) - 4m^4(D-1)$. We notice that once $\Box-\slashed{\p}^2=0$ in any dimension, the theory propagates a spin-3/2 particle for arbitrary values of $D$. In the special case of $D=3$ that we are interested here, we have the following expression for the propagator:
\be
G^{-1}_{\mu\nu} = \frac{1}{2(\Box - m^2)}\left(P^{(3/2)}\right)_{\mu\nu} + \frac{1}{4m^2}\left(P^{(1/2)}\right)_{\mu\nu} - \frac{1}{2\sqrt{2} m^2}\left(P^{(1/2)}_{12} + P^{(1/2)}_{21}\right)_{\mu\nu}.
\ee

\no Using conserved sources, and writing the propagator in the momentum space we have the transition amplitude saturated between sources $\bar{j}^{\mu}$ and $j_{\mu}$:

\bea
\mathcal{A}_2(p) &=& - \frac{i}{2} \bar{j}^\mu G_{\mu\nu}^{-1}j^\nu \\ \nonumber
&=& - \frac{i}{4}\bar{j}^\mu \left[- \frac{1}{p^2 + m^2}\left(P^{(3/2)}\right)_{\mu\nu} + \frac{1}{2m^2}\left(P^{(1/2)}_{22}\right)_{\mu\nu} - \frac{1}{\sqrt{2}m^2}\left(P^{(1/2)}_{12} + P^{(1/2)}_{21}\right)_{\mu\nu}\right]j^\nu.
\eea

The imaginary part of the residue of the transition amplitude is then given by:

\be
Im\left[Res\left(\mathcal{A}_2(p)\right)\right]|_{p^2 = -m^2} = \frac{1}{4}\bar{j}^{\mu}\left(P^{(3/2)}\right)_{\mu\nu}j^\nu= \frac{1}{4}\bar{j}_i j_i > 0.
\ee

\no Where we have used the convenient rest frame $k_{\mu}=(m,0,...,0)$ and the fact that the source must be conserved $\partial_\mu j^\mu = \partial_\mu \bar{j}^\mu = 0$ and gamma-traceless $
\gamma_\mu j^\mu = \gamma_\mu \bar{j}^\mu = 0$ \footnote{Notice that, in the original works by \cite{deserkay,deser3/2} the authors have defined the massive spin-3/2 lagrangians with a gobal complex factor $i$, which is necessary in order to have a hermitian lagrangian. Such global factor has been neglected in our master actions because it does not interfere in the results. However one have to notice that, in the signature and representation of the gamma matrices used here $\bar{j}_{\mu}=j_{\mu}^{\dagger}\gamma^{0}$ and $\gamma^{0\dagger}=-\gamma^{0}$. Which implies $(\bar{j}_{\mu}j^{\mu})^{\dagger}=-\bar{j}_{\mu}j^{\mu}$. So, in order to have a hermitian result one needs to take in account the complex global factor.}. Once we have such result we can conclude that the second order doublet model suggested in \cite{nge32} is free of ghosts containing a unique massive spin-3/2 doublet in its spectrum. Through the master action technique we will see that it is quantum equivalent to a fourth order doublet model, similar to the spin-2 linearized  NMG model.

\section{Interpolating among self-dual descriptions}

There exists in the literature three self-dual models describing massive spin-3/2 particles in $D=2+1$. The second order action given by (\ref{sd2}) was initially introduced by \cite{deserkay}. Subsequently it was demonstrated in $\cite{deser3/2}$ that it is indeed equivalent, through the introduction of a master action, to the first order self-dual model (\ref{sd1}). By using symmetry arguments, in what we call the Noether Gauge Embedment technique, some of us in \cite{nge32} have demonstrated that it is possible to obtain a third-order self-dual model from the second-order one, such that we have a sequence of three self-dual descriptions connected by dual relations at the classical level. It turns out that such third order description is the fermionic part of the Topologically Massive Supergravity theory suggested by \cite{Andringa}. However, up to now we do not have any guarantee that the three self-dual models are equivalent at the quantum level. In order to establish the quantum equivalence we now suggest a unique triple master action interpolating among the three self-dual descriptions. It is given by: 

\bea S_M&=& -\int \, \bar{\psi}\,\cdot d \psi + m\int \, (\bar{\psi}\, \psi)+ \int \, (\bar{\psi}-\bar{\chi})\,\cdot d (\psi-
\chi) + \frac{1}{2m} \int \, (\bar{\phi}-\bar{\chi})\,\cdot d \Omega(\phi-
\chi).\nn\\ \label{SM} \eea

As we have discussed in \cite{dualdesc} the key ingredient to construct master actions is the introduction of mixing terms free of particle content (the third and the fourth terms respectively of the action given by (\ref{SM})). Here we have introduced two of them, the Chern-Simons-like term and the second order one. Finally, we have to say that  now we have two auxiliary fields $\chi$ and $\phi$.

In order to compare correlation functions we now introduce a generating functional by adding to the master action source terms $j_{\mu}$ and $\bar{j}_{\mu}$, then we have:

\be W[\bar{j},j]=\int \, {\cal{ D}}\bar{\psi}\, {\cal{ D}}{\psi}\, {\cal{ D}}\bar{\chi}\, {\cal{ D}}{\chi}\, {\cal{ D}}\bar{\phi}\, {\cal{ D}}{\phi} \exp i\left[ S_M+ \int\, d^3x\, (\bar{j}^{\mu}\psi_{\mu} + \bar{\psi}^{\mu}j_{\mu})\right].\label{WM}\ee

It is straightforward to check that performing the trivial shifts $\bar{\phi}\to \bar{\phi}+\bar{\chi}$, ${\phi}\to {\phi}+{\chi}$ and then $\bar{\chi}\to \bar{\chi}+\bar{\psi}$, ${\chi}\to {\chi}+{\psi}$ in (\ref{WM}),  the mixing terms get completely decoupled, and once they are free of particle content one can functionally integrate out both of them which give us:

\be W[\bar{j},j]=\int \, {\cal{ D}}\bar{\psi}\, {\cal{ D}}{\psi}\, \exp i\left[ S_{SD(1)}+ \int\, d^3x\, (\bar{j}^{\mu}\psi_{\mu} + \bar{\psi}^{\mu}j_{\mu})\right].\label{Wsd1}\ee

\no In this way one can now obtain correlation functions deriving with respect to the sources in (\ref{WM}) and (\ref{Wsd1}) in order to demonstrate that the first order self-dual model is equivalent to the master action. 

\be \langle \psi_{\mu 1}(x_1) \, ...\, \psi_{\mu N}(x_N)\rangle_{M} =  \langle \psi_{\mu 1}(x_1) \, ...\, \psi_{\mu N}(x_N)\rangle_{SD(1)},\ee
\be \langle \bar{\psi}_{\mu 1}(x_1) \, ...\, \bar{\psi}_{\mu N}(x_N)\rangle_{M} =  \langle \bar{\psi}_{\mu 1}(x_1) \, ...\, \bar{\psi}_{\mu N}(x_N)\rangle_{SD(1)}.\ee

However, in (\ref{SM}) if we work with the third term opening it we can rearrange the master action considering the source terms in the following way:

\bea S_M&=&  m\int \, (\bar{\psi}\, \psi)+ \int \,d^3x \left( \bar{\psi}_{\mu} {\cal U}^{\mu} +\bar{{\cal U}}^{\mu}\psi_{\mu}\right)+  \frac{1}{2m} \int \, (\bar{\phi}-\bar{\chi})\,\cdot d \Omega(\phi-
\chi),\nn\\ \label{SM2} \eea

\no where we have defined  $\bar{{\cal U}}^{\mu}= \bar{f}^{\mu}(\chi)+\bar{j}^{\mu}$ and ${{\cal U}}^{\mu}= {f}^{\mu}(\chi)+{j}^{\mu}$. One can now notice that we can completely decouple $\psi$ implementing the shifts $\bar{\psi}_{\mu}\to \bar{\psi}_{\mu}-\bar{{\cal U}}^{\nu}\gamma_{\mu}\gamma_{\nu}/2m$ and ${\psi}_{\mu}\to {\psi}_{\mu}-\gamma_{\nu}\gamma_{\mu}{{\cal U}}^{\nu}/2m$ in $(\ref{SM2})$ which gives rise to the decoupled mass term $m\int\, (m^2)$ free of particle content. Back in the generating functional (\ref{WM}) and integrating over $\psi$ we have:

\bea W[\bar{j},j]&=&\int \,  {\cal{ D}}\bar{\chi}\, {\cal{ D}}{\chi}\, {\cal{ D}}\bar{\phi}\, {\cal{ D}}{\phi} \exp i\left[ S_{SD(2)}+ \frac{1}{2m} \int \, (\bar{\phi}-\bar{\chi})\,\cdot d \Omega(\phi-
\chi)\right.\nn\\
&+&\left. \int\, d^3x\,\left (\bar{j}^{\mu}F_{\mu}(\chi) + \bar{F}^{\mu}(\chi)j_{\mu}+\frac{1}{2m} \bar{j}^{\mu}\gamma_{\nu}\gamma_{\mu}j^{\nu}\right)\right].\label{WM3}\eea

\no It is straightforward to see that the shifts $\bar{\phi} \to \bar{\phi}+\bar{\chi}$ and ${\phi} \to {\phi}+{\chi}$ makes the remaining mixing term in (\ref{WM3}) decoupled, once it is free of particle content, after integrating over $\phi$ and then deriving with respect to the sources one can conclude:

\be \langle \psi_{\mu 1}(x_1) \, ...\, \psi_{\mu N}(x_N)\rangle_{SD(1)} = \langle F_{\mu 1}(\chi) \, ...\, F_{\mu N}(\chi)\rangle _{SD(2)}\quad +\quad  C.T , \ee

\be  \langle \bar{\psi}_{\mu 1}(x_1) \, ...\, \bar{\psi}_{\mu N}(x_N)\rangle_{SD(1)} =\langle \bar{F}_{\mu 1}(\chi) \, ...\,\bar{F}_{\mu N}(\chi)\rangle _{SD(2)} \quad +\quad  C.T , \ee		

\no where $C.T$ stands for contact terms coming from the  quadratic terms on the sources, they are proportional to $\sim \delta_{\mu\nu}\delta(x-y)$, then for the purposes of this work they are not important at all. Besides we have obtained the dual maps $F_{\mu}$ which are gauge invariant and  precisely coincides with the one we have obtained at classical level in \cite{nge32}:

\be {\psi}_\nu \leftrightarrow	{F}_{\nu}= \frac{{\Omega}_{\nu}(\chi)}{2m}\quad; \quad  \bar {\psi}_\nu \leftrightarrow	\bar{F}_{\nu}= \frac{\bar{\Omega}_{\nu}(\chi)}{2m}\ee

Finally one can interpolate the generating functional in order to obtain the third order self-dual model. To make it possible, we get back to the expression (\ref{WM3}) (without any shifts) and work with the mixing term taking in account the important property:

\be \int \, \bar{\phi} \cdot d \Omega(\chi) = \int \, \bar{\Omega}(\phi)\cdot d \chi.\ee

\no With this result in hand we can rewrite the master action from (\ref{WM3}) in such a way that we have:
 
 \be S_M= \int\, \left(\bar{\chi}-\frac{\bar{\Omega}(\phi)}{2m}\right) \cdot d \left(\chi-\frac{\Omega(\phi)}{2m}\right)-\frac{1}{4m^2}\int \, \bar{\Omega}(\phi)\cdot d\Omega(\phi)+ \frac{1}{2m}\int \,\bar{\phi}\cdot d \Omega(\phi).\label{SM3}\ee

\no Plugging back (\ref{SM3}) in (\ref{WM3}) one can notice that with the shifts $\bar{\chi}\to \bar{\chi} + \bar{\Omega}(\chi)/2m$ and $\chi \to \chi +\Omega(\chi)/2m$ the fields $\chi $ and $\phi$ get decoupled. In order to decouple the fields $\chi$ from the sources we then implement $\bar{\chi}_{\mu}\to \bar{\chi}_{\mu}-\bar{j}^{\nu}\gamma_{\mu}\gamma_{\nu}/2m$ and ${\chi}_{\mu}\to {\chi}_{\mu}-{j}^{\nu}\gamma_{\mu}\gamma_{\nu}/2m$. After all these shifts we end up with a decoupled Chern-Simons like term on $\chi$ which is free of particle content. One can then integrate over $\chi$ to have:

\bea W[\bar{j},j]&=&\int \,  {\cal{ D}}\bar{\phi}\, {\cal{ D}}{\phi} \exp i\left\lbrace S_{SD(3)}+ \int\, d^3x\,\left[\bar{j}^{\mu}F_{\mu}\left(\frac{\Omega(\phi)}{2m}\right) + \bar{F}^{\mu}\left(\frac{\Omega(\phi)}{2m}\right)j_{\mu}+\frac{1}{2m} \bar{j}^{\mu}\gamma_{\nu}\gamma_{\mu}j^{\nu}\right]\right\rbrace\nn\\ \label{WM4}\eea

Deriving with respect to the sources in (\ref{WM3}) and (\ref{WM4}) one concludes on the quantum equivalence between the third and the second order self-dual models, which is given by the comparison of the correlation functions:

\be \langle F_{\mu 1}(\chi) \, ...\, F_{\mu N}(\chi)\rangle _{SD(2)} = \left\langle \bar{F}_{\mu_1}\left(\frac{\Omega(\phi)}{2m}\right) \, ...\,\bar{F}_{\mu_N}\left(\frac{\Omega(\phi)}{2m}\right)\right\rangle _{SD(3)} \quad +\quad  C.T , \ee

\be \langle \bar{F}_{\mu 1}(\chi) \, ...\,\bar{F}_{\mu N}(\chi)\rangle _{SD(2)}=\left\langle {F}_{\mu_1}\left(\frac{\Omega(\phi)}{2m}\right) \, ...\,{F}_{\mu_N}\left(\frac{\Omega(\phi)}{2m}\right)\right\rangle _{SD(3)} \quad +\quad  C.T.  \ee

The entire third order self-dual action with the sources included is invariant under the local transformations $\delta \phi_{\mu} = \p_{\mu}\Lambda + \gamma_{\mu}\xi$ and $\delta \bar{\phi}_{\mu} = \p_{\mu}\bar{\Lambda} + \bar{\xi}\gamma_{\mu}$. The results are the same we have obtained in \cite{nge32} and one can say that the equivalence among the three self-dual descriptions can perfectly be extended to the quantum level.

\section{Interpolating between doublet models}
Let us introduce the following master action:

\be S_{M}= \int \,\left[ -\frac{1}{4}\bar{\psi}\,\cdot d \Omega(\psi) + \frac{m^2}{2} \, (\bar{\psi}\, \psi)+\frac{1}{4}(\bar{\psi}-\bar{\chi})\,\cdot d \Omega(\psi-\chi) \right].\label{SMD}\ee

\no An important consideration must be made now. Notice that, in order to construct the master action one takes the second order doublet model (\ref{d2}) and then add to it  a mixing term of second order in derivatives. The reader can associate the second order model to the spin-1 Proca theory, and in analogy one would add the Maxwell term to construct a spin-1 master action, but this would take us inevitably to the fourth order Podolski model which we know contains ghost in its spectrum. Differently if in analogy one starts with the spin-2 linearized Fierz-Pauli theory and then add to it the second order Einstein-Hilbert term to construct a master action, such action would interpolate between the first theory and a fourth order theory which we recognize as being the linearized version of the New Massive Gravity free of ghosts. The key ingredient here is the difference of mixing terms, once in the spin-1 case the Maxwell term is not free of particle content in $D=2+1$, one can construct a master action but the fourth order theory would contain ghosts,  while in the spin-2 case the linearized Einstein-Hilbert term is trivial in $D=2+1$ which provides us a good fourth order model \cite{nsd}. 

In this sense, once we know through the unitarity analysis of the second order model that it is free of ghosts, and once we know that the second order mixing term is free of particle content, we expect the master action (\ref{SMD}) interpolating between good second and fourth order doublet models. In order to do that we add source terms putting the action on a generating functional:

\be W[\bar{j},j]=\int \, {\cal{ D}}\bar{\psi}\, {\cal{ D}}{\psi}\, {\cal{ D}}\bar{\chi}\, {\cal{ D}}{\chi}\, \exp i\left[ S_M+ \int\, d^3x\, (\bar{j}^{\mu}\psi_{\mu} + \bar{\psi}^{\mu}j_{\mu})\right].\label{WM2}\ee

\no It is trivial to see that the shifts $\bar{\chi}\to \bar{\chi}+\bar{\psi}$ and ${\chi}\to {\chi}+{\psi}$ completely decouple the mixing term allowing us to functionally integrate over the fields $\chi$, which would lead us to the conclusion that the master action is equivalent to the second order doublet model by comparing its correlation functions. On the other hand if work with the mixing term  the master action considering source terms becomes:

\be S_{M}= \int \, \frac{m^2}{2} \, (\bar{\psi}\, \psi) + \int \,d^3x \left( \bar{\psi}_{\mu} {\cal V}^{\mu} +\bar{{\cal V}}^{\mu}\psi_{\mu}\right)+\frac{1}{4}\int \,\bar{\chi}\,\cdot d \Omega(\chi)  .\label{SMD2}\ee

\no Where we have defined $\bar{{\cal V}}^{\mu}= -\epsilon^{\mu\nu\alpha}\p_{\nu}\bar{\Omega}_{\alpha}/4+\bar{j}^{\mu}$  and ${{\cal V}}^{\mu}= -\epsilon^{\mu\nu\alpha}\p_{\nu}{\Omega}_{\alpha}/4+{j}^{\mu}$. Then exactly as before we know that the decoupling of the fields $\psi$ from ${\cal{V}}$ can be reached through the transformations $\bar{\psi}_{\mu}\to \bar{\psi}_{\mu}-\bar{{\cal V}}^{\nu}\gamma_{\mu}\gamma_{\nu}/m^2$ and ${\psi}_{\mu}\to {\psi}_{\mu}-\gamma_{\nu}\gamma_{\mu}{{\cal V}}^{\nu}/m^2$. Back with (\ref{SMD2}) in the generating functional (\ref{WM2}), after an integration of a decoupled mass term on $\psi$ we have:

\be W[\bar{j},j]=\int\, {\cal{ D}}\bar{\chi}\, {\cal{ D}}{\chi}\, \exp i\left\lbrace S_{D(4)}+ \int\, d^3x\, \left[\bar{j}^{\mu}G_{\mu}(\chi) + \bar{G}^{\mu}(\chi)j_{\mu}+{\cal{O}}(j^2)\right]\right\rbrace,\label{WMD4}\ee

\no where ${\cal O}(j^2)$ refers to terms quadratic on the sources. The dual maps in this case are given by the spinor vectors $G_{\mu}(\chi)$ which are invariant under the local transformations of the type reparametrizations and gamma-Weyl; they are explicitly given by:

\be \bar{\psi}_{\mu}\leftrightarrow \bar{G}_{\mu}= \frac{\bar{\Omega}_{\mu}(\Omega(\chi))}{4m^2}\quad;\quad {\psi}_{\mu}\leftrightarrow G_{\mu}=\frac{{\Omega}_{\mu}(\Omega(\chi))}{4m^2}.\ee

\no The dual maps we have obtained here are exactly the same we have found in \cite{nge32}. They provide an interpolation of the equations of motion as well as they establish the equivalence of the correlation functions if we derive with respect to the source terms on the generating functional i.e.:

\be \langle \psi_{\mu 1}(x_1) \, ...\, \psi_{\mu N}(x_N)\rangle_{D(2)} = \langle G_{\mu 1}(\chi) \, ...\, G_{\mu N}(\chi)\rangle _{D(4)}\quad +\quad  C.T , \ee

\be  \langle \bar{\psi}_{\mu 1}(x_1) \, ...\, \bar{\psi}_{\mu N}(x_N)\rangle_{D(2)} =\langle \bar{G}_{\mu 1}(\chi) \, ...\,\bar{G}_{\mu N}(\chi)\rangle _{D(4)} \quad +\quad  C.T . \ee		

Notice that once we have quadratic terms on the source the equivalence will be satisfied in the presence of contact terms. Besides, one can check that the fourth order theory, as well as the dual maps, are invariant under reparametrizations while the fourth order term is invariant under gamma-Weyl transformation much like what happens with the spin-2 examples of the Fierz-Pauli theory and the NMG theory.

\section{Conclusion}
A triple master action is suggested in order to connect three self-dual descriptions of massive spin-3/2 particles in $D=2+1$ dimensions. This is a generalization of that first introduced by Deser and Kay in \cite{deserkay}. Here we use the idea of considering trivial mixing terms like the Chern-Simons $\bar{\psi} \cdot d \psi$ and the second order term $\bar{\psi} \cdot  d \Omega(\psi)$ to construct our triple master action and then through the addition of source terms we define a generating functional which allow us the comparison of correlations functions. It is interesting to notice that a shorthand notation used in the bosonic cases of models describing spin-2 and spin-3 particles, can also be used in the fermionic case, as we have defined in section-2.

In \cite{soldering32}, we have demonstrated that the first order self-dual model of \cite{deserkay} can be obtained by a square-root of a second order doublet model. In addition, by soldering second order self-dual models with different helicities $+3/2$ and $-3/2$ we obtain a fourth order doublet model describing both helicities. Here we suggest a master action which interpolates between the doublet descriptions of second and fourth order in derivatives showing that they are indeed quantum equivalents by comparing correlation functions. The fourth order model has several similarities with the linearized spin-2 counterpart which is named the linearized NMG model. The construction of the present master action interpolating between models free of ghosts is possible thanks to the triviality of the second order term as discussed before. This also happens in the spin-2 case, while it is not possible in the spin-1 case.

In order to check the particle content of the second order doublet model (\ref{d2}) we use a spin-projection basis to put the lagrangian in a bilinear form and then to find the propagator. We have verified that the model correctly describes a massive spin-3/2 doublet and that the model is unitary in $D$ dimensions. Notice that, however the fourth-order model is free of ghosts in three dimensions, this is not a guarantee that the fourth order doublet model is also unitary in $D$ dimensions, once the construction of the master action is done taking the fact that the second-order mixing term is trivial just in three dimensions. Master actions proofing the equivalence of pairs of singlets and doublets and also an analysis of unitarity of the fourth order doublet model in $D$ dimensions are in course.

\section{Acknowledgements}

H.L.O and P.H.F.N  are supported by CAPES. The authors thanks to Gustavo Pazzini de Brito and Alessandro Ribeiro, for discussions.

\end{document}